\documentclass[pra,twocolumn,showpacs,amsmath,amssymb,superscriptaddress]{revtex4}
\usepackage{epsfig}
\usepackage{color}
\usepackage{graphicx}

\emergencystretch=\maxdimen
\begin{document}

\title{Exotic phases of interacting $p$-band bosons}
\author{F. H\'ebert}
\affiliation{
INLN, Universit\'e de Nice-Sophia Antipolis, CNRS;
1361 route des Lucioles, 06560 Valbonne, France.
}
\author{Zi Cai}
\affiliation{
Physics Department, Arnold Sommerfeld Center for Theoretical
  Physics, and Center for NanoScience,
  Ludwig-Maximilians-Universit\"at M\"unchen, D-80333 M\"unchen,
  Germany.
}
\affiliation{
Department of Physics, University of California, San Diego, CA
  92093, USA. 
}
\author{V. G. Rousseau}
\affiliation{
Department of Physics and Astronomy, Louisiana State University,
  B\^aton Rouge, Louisiana 70803, USA.
}
\author{Congjun Wu}
\affiliation{
Department of Physics, University of California, San Diego, CA
  92093, USA.
}
\author{R. T. Scalettar}
\affiliation{
Physics Department, University of California, Davis, CA 95616,  USA.
}
\author{G. G. Batrouni}
\affiliation{
INLN, Universit\'e de Nice-Sophia Antipolis, CNRS;
1361 route des Lucioles, 06560 Valbonne, France.
}
\affiliation{
Institut Universitaire de France
}

\begin{abstract}
  We study a model of interacting bosons that occupy the first excited
  $p$-band states of a two-dimensional optical lattice.  In contrast
  to the much studied single band Bose-Hubbard Hamiltonian, this more
  complex model allows for non-trivial superfluid phases associated
  with condensation at non-zero momentum and staggered order of the
  orbital angular momentum in addition to the superfluid-Mott
  insulator transition.  More specifically, we observe staggered
  orbital angular momentum order in the Mott phase at commensurate
  filling and superfluidity at all densities.  We also observe a
  transition between the staggered angular momentum superfluid phase
  and a striped superfluid, with an alternation of the phase of the
  superfluid along one direction. The transition between these two
  phases was observed in a recent experiment, which is then
  qualitatively well described by our model.
\end{abstract}

\maketitle

\section{Introduction}

Superfluidity has attracted much attention since its discovery in
bosonic $^4$He and, later, in fermionic $^3$He \cite{tilley90}.  This
phenomenon was studied in a wide range of systems ranging from
excitons in quantum wells \cite{mosalenko62} to neutron stars
\cite{migdal59}. Interest intensified following the experimental
realization of confined ultracold atomic systems, in particular atomic
Bose-Einstein condensates (BEC) \cite{zwierlein05}.  Many new
possibilities become available when these systems are loaded in the
lowest band of optical lattices where they are governed by the
(bosonic or fermionic) Hubbard model with highly tunable parameters
\cite{jaksch98}. After their initial use to
  explore quantum phase transition between superfluid (SF) and Mott
  insulator (MI) phases \cite{greiner02} 
  ultracold atomic gases have been used to study mixtures of particles
\cite{zwierlein06,liao10} 
and, since then, more exotic
  pairing phenomena, including Fulde-Ferrell-Larkin-Ovchinnikov
  \cite{fulde64,larkin65} or breached paired phases
  \cite{sarma63,liu03} in unbalanced fermionic systems, or to
  introduce the concept of counter-superfluidity in MI of boson
  mixtures \cite{kuklov03}.  Work on spinor condensates concentrated
on the interplay between superfluid behavior and itinerant magnetism,
especially through the study of spin-1 bosons
\cite{stamper01,kawaguchi12}.

More recently, it was proposed to load the atoms in higher bands of
the optical lattice in order to study further exotic forms of
superfluidity \cite{isacsson05}.  In a three dimensional cubic
lattice, there are three such states that are degenerate and
correspond to the different states of orbital angular momentum $l=1$
(${\bf L}^2 = l(l+1) = 2$ in units of $\hbar^2$), whereas the unique
ground state corresponds to $l=0$.  Due to the anisotropy used to produce
two dimensional square lattices, one of the orbitals has a larger
energy than the other two.  This reduces the model to only two species
(the $p_x$ and $p_y$ states) in the low energy limit.  On cubic or
square lattices, the hopping parameters from a site to its neighbors
are anisotropic and take two very different values depending on the
hopping direction, parallel or perpendicular to the orbital axis.
Isacsson and Girvin \cite{isacsson05} studied the limit where the
hopping in the transverse directions is totally suppressed. This led
them to suggest that in two dimensions their model may develop a
peculiar columnar phase ordering where the phases of particles in the
$p_{x(y)}$ states are coherent along the $x(y)$ direction and
uncorrelated in the transverse direction.

Bosons in high orbital bands are not in the true ground state. This
feature gives rise to the possibility of new states of matter beyond
the ``no-node'' theorem \cite{feynman98} that is obeyed by the
conventional BECs of single component bosons.  This theorem states
that the many-body ground state wavefunctions of bosons under very
general conditions are positive-definite.  It implies that
time-reversal symmetry cannot be spontaneously broken.  If the system
has rotational symmetry, this theorem constrains the condensate
wavefunctions to be rotationally invariant.  In other words,
conventional BECs are $s$-wave-like whose symmetry property is similar
to $s$-wave superconductivity.  Recently, unconventional symmetries
have been introduced to the single-boson condensates in high orbital
bands in optical lattices \cite{wu09,liu06}, denoted as
``unconventional BECs'' (UBECs).  Their condensate wave functions
belong to nontrivial representations of the lattice point group.  In
other words, they are non-$s$-wave in analogy to unconventional
pairing symmetries of superconductivity.  Liu and Wu \cite{liu06}
studied analytically the UBECs in the $p$-orbital band with non zero
transverse hopping exhibiting a $p_x\pm ip_y$ type symmetry, and thus
breaking time-reversal symmetry spontanously with a complex-valued
condensate wavefunction.  They predicted for densities, $\rho$, larger
than two particles per site, the existence of a superfluid phase where
the system condenses at non zero quasi-momentum and which is
accompanied by a staggered order of the orbital angular momentum
(SAM). Recently, the model was studied using an effective action approach \cite{li11} and a similar SF phase
with SAM order was predicted even for $\rho=1$ as well as an antiferromagnetic Mott phase. Its physics 
was also examined in one dimension \cite{li12}.  For unconventional symmetries such as the $d$-wave of high
T$_c$ cuprates, phase-sensitive detections provide definitive evidence
\cite{wollman1993,tsuei1994}.  For the non-$s$-wave UBECs,
phase-sensitive detection on condensate symmetries has also been
proposed through Raman transition \cite{cai2012}.

Such an exotic SF phase was observed in a recent experiment by Wirth,
\"Olschl\"ager, and Hemmerich \cite{wirth11,olschlager12} in a two
dimensional checkerboard lattice composed of $s$ and $p$-orbitals
sites and had been investigated theoretically by Cai et
al. \cite{cai11}.  The nearest neighbors of $s$-sites being $p$-sites,
an atom cannot go directly from a $p$-site to another as was the case
in Liu and Wu's model \cite{liu06,wu09}. However, the $s$-sites do not
introduce phase differences and thus only play the r\^ole of a neutral
relay between $p$-sites.  Introducing a small anisotropy between the
$x$ and $y$ axes, a transition between the condensed state at non zero
momentum and a striped phase where there is a phase alternation
between different stripes was observed.  BECs with unconventional
condensate symmetries have also been observed in the solid state
exciton-polariton lattice systems \cite{kim2011}.

In this work, we will study the model originally proposed by Liu and
Wu \cite{liu06} with quantum Monte Carlo simulations. To reproduce
qualitatively the different phases observed in Hemmerich's group
experiment, we will add to the original model an anisotropy term in
the form of an energy difference between $p_x$ and $p_y$
orbitals. Finally, in addition to studying the superfluid phase, we
will also focus on the insulating phases that arise in such systems
for strong enough interactions and integer densities.  In section II
of this article, we will introduce and discuss the model, in section
III we will show that it can be mapped on a bosonic spin-1/2 model and
establish the correspondence between the phases observed.  In section
IV, we will present the results of our simulations and conclude in
section V.

\section{The $P$-Orbital model}
\begin{figure}
\includegraphics[width=8.5cm]{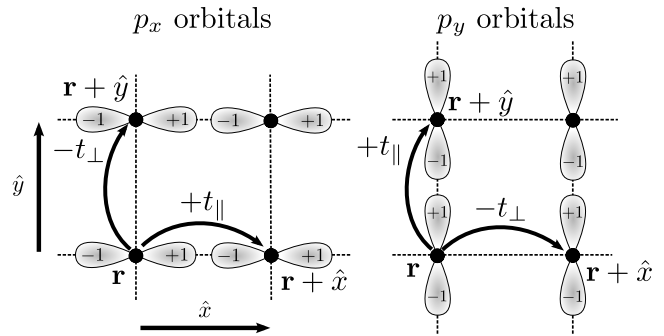}
\caption{Hopping parameters on the square lattice for $p_x$ and $p_y$ 
orbitals.  Because $p_{x,(y)}$ orbitals are parity odd,
the overlap of orbitals changes sign depending on the direction. 
Along the parallel  direction ($x$ for $p_x$ orbitals, $y$ for $p_y$ orbitals), 
the overlap is negative, which gives a positive $+t_\parallel$ hopping 
parameter. On the other hand, along the perpendicular  direction, 
there is a conventional negative hopping parameter $-t_\perp$.
\label{orbitals}}
\end{figure}

We will study the model introduced in \cite{liu06,wu09} for the two-dimensional
square lattice and two-species (``spin-$\frac{1}{2}$") case. 
The system is then governed by the Hamiltonian,
\begin{eqnarray}
\nonumber
H &=& +t_\parallel \sum_{\bf r}\left(p^\dagger_{x, \bf r} 
p^{\phantom{\dagger}}_{x, {\bf r}+\hat{x}}  
+ p^\dagger_{y, \bf r} 
p^{\phantom{\dagger}}_{y, {\bf r}+\hat{y}} + {\rm
  h.c.}\right) \nonumber\\ 
\nonumber
&&- t_\perp  \sum_{\bf r} \left(p^\dagger_{x, \bf r}
p^{\phantom{\dagger}}_{x, {\bf
    r}+\hat{y}}  + p^\dagger_{y, \bf r} 
p^{\phantom{\dagger}}_{y, {\bf r}+\hat{x}} + {\rm
  h.c.}\right) \label{hopping}\\ 
&&+\frac{U}{2} \sum_{\bf r} \left(n_{\bf r}^2 - \frac{L_{z, {\bf r}}^2}{3} \right) 
+ \Delta \sum_{\bf r}\left(n_{x,{\bf r}} -
n_{y,{\bf r}}\right), 
\label{interaction} 
\end{eqnarray}
where $p_{x(y),{\bf r}}$ is the destruction operator of a particle
located on site ${\bf r} = (r_x,r_y)$ of an $L\times L$ square lattice
in the $p_x (p_y)$ state; $\hat x$ and $\hat y$ are the primitive
vectors of the square lattice.  The number operators are $n_{x(y),{\bf
    r}} = p^\dagger_{x(y),{\bf r}} p^{\phantom{\dagger}}_{x(y),{\bf
    r}}$ and $n_{\bf r}= n_{x,{\bf r}} + n_{y,{\bf r}}$; and $L_z$ is
the on site orbital angular momentum operator defined as
\begin{eqnarray}
L_{z,{\bf r}} &=& -i (p^\dagger_{x, \bf r} p^{\phantom{\dagger}}_{y, {\bf r}} -
p^\dagger_{y,\bf r} p^{\phantom{\dagger}}_{x, {\bf r}}),
\label{Lz} 
\end{eqnarray}
We remark that $L_{z,{\bf r}}^2$ is not diagonal in this basis and
contains terms that transform two particles of one species into two
particles of the other species.

Since the overlap of $p$-orbitals on neighboring sites is different in
the directions parallel or perpendicular to their spatial orientation,
there are two different hopping parameters.  $t_\perp$ is typically
smaller than $t_\parallel$ (Fig. \ref{orbitals}).  Moreover, due to
the phase difference between the two parts of the $p$-orbitals, the
two hopping terms have different signs (Eq. (\ref{hopping})), being
positive in the parallel direction ($+t_\parallel$) whereas the
perpendicular hopping term maintains the conventional negative sign
($-t_\perp$).  We will concentrate on the case where
$t_\perp=t_\parallel=t$, but also retain the sign difference.  The
parameter $t$ sets the energy scale.  The interaction part of the
Hamiltonian (Eq. (\ref{interaction})) includes a conventional on-site
repulsion (the $n^2_{\bf r}$ term) and a term that maximizes the
on-site angular momentum (the $L^2_{z,{\bf r}}$ term).  This is
essentially the physics of second Hund's rule applied to the bosonic
orbital system: complex orbitals are spatially more extended to save
repulsive interactions.  The last term in the Hamiltonian,
Eq. (\ref{interaction}) corresponds to a tunable difference in energy,
$\Delta$, between the $p_x$ and $p_y$ orbitals, due to a corresponding
anisotropy in the lattice.

In the non interacting limit, the energy dispersion for $p_x$
particles \cite{liu06} takes the form $\epsilon_{x}({\bf
  k})=2t(\cos(k_x) - \cos(k_y))$ where ${\bf k} = (k_x,k_y)$,
$k_{x(y)} = 2\pi K_{x(y)}/L$, and $K_{x(y)}$ is an integer. For $p_y$
particles $\epsilon_{y}({\bf k}) = - \epsilon_{x}({\bf k})$.  Since
the energy minima appear at ${\bf k}=(\pi,0)$ for the $p_x$ particles
and at ${\bf k}=(0,\pi)$ for the $p_y$ particles, it is expected for
the system to condense at non zero momentum.

\begin{figure}[h]
\includegraphics[width=7cm]{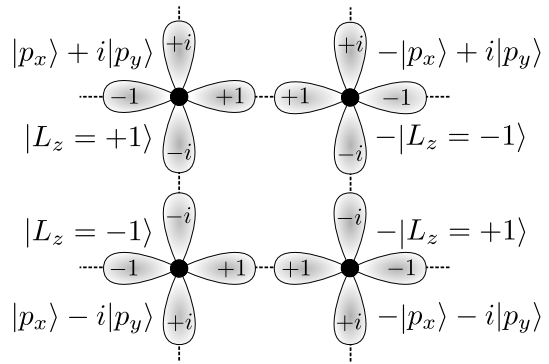}
\caption{Staggered orbital angular momentum order. 
  A configuration having states proportional to $|L_z = \pm 1\rangle$
  on each site maximizes $L_z^2$, minimizing the interaction
  energy. Concentrating on the $p_x$ orbital, there is a phase
  alternation along the $x$ direction and a phase coherence along the
  $y$ direction, which minimizes the kinetic energy and corresponds to
  the condensation with wave vector ${\bf k} = (\pi,0)$. The same
  phenomenon is observed for the $p_y$ orbitals, with reversed axes and
  ${\bf k}=(0,\pi)$. This gives an alternation of sites with $L_z= \pm
  1$ and thus a staggered order for the angular momentum along $z$.
\label{stagorbitals}
}
\end{figure}

In the interacting case, the interaction energy is minimized by
maximising $L^2_{z,{\bf r}}$ on a given site, that is by putting all
the particles in the same state corresponding to either $L_z = +1$ or
$L_z=-1$: $|L_z = +1\rangle \propto |p_x\rangle + i|p_y\rangle $ and
$|L_z = -1\rangle \propto |p_x\rangle -i |p_y\rangle$.  With these
on-site states, it is then possible to minimize the hopping energy by
using a configuration \cite{liu06} that gives a phase difference along
the longitudinal hopping and a phase match for transverse
hopping. Such a configuration is represented in
Fig.~\ref{stagorbitals}; it exhibits a checkerboard pattern of $L_z =
\pm 1$ sites. This is the staggered angular momentum order mentioned
in the introduction, and it is clear that it is compatible with a
phase ordering that corresponds to the condensation at non zero
momentum for both $p_x$ and $p_y$ particles.

\begin{figure}
\includegraphics[width=7cm]{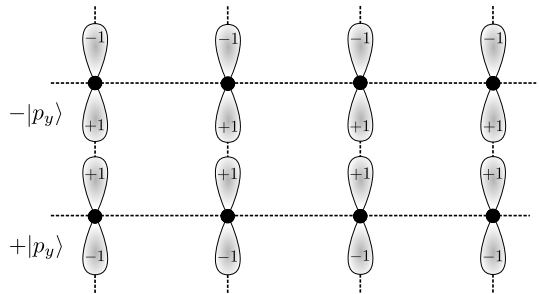}
\caption{Stripe phase. Due to the different signs of
  the hopping parameters, when the system is composed of just one
  species ($p_y$ in this figure) it will have phase alternation along
  the parallel ($y$) direction and phase coherence along the
  transverse ($x$) direction to minimize the kinetic energy.  This
  forms stripes aligned with the transverse direction and striped
  superfluid phases can be observed.
\label{stripes}}
\end{figure}

Finally the $\Delta$ term in Eq. (\ref{interaction}) should suppress
this kind of order since it requires having the same number of $p_x$
and $p_y$ particles by increasing the density of $p_y$ particles. If
the system remains superfluid when it is composed mostly of $p_y$
particles, the phase will alternate between sites along the $y$
direction and will be coherent in the $x$ direction, thus forming
stripes along the $x$ axis. We will call this phase a striped
superfluid (see Fig.~\ref{stripes}).

\section{Mapping on a spin 1/2 model}

The positive parallel hopping term in Eq.~\ref{hopping} generates a
sign problem for Quantum Monte Carlo simulations.  (In this work we
use the Stochastic Green Function algorithm
\cite{rousseau08,rousseau08-2}).  However, the Hamiltonian can be
mapped onto a spin$-1/2$ model free of this problem
\cite{deforges11,deforges12}.  We define the spin 1/2 bosonic
operators
\begin{equation}
b_{\uparrow,{\bf r}} = i (-1)^x p_{x,{\bf r}}
\quad
b_{\downarrow,{\bf r}} =  (-1)^y p_{y,{\bf r}}
\end{equation}
This transformation gives a direct equivalence between densities of
$\uparrow (\downarrow)$ and $p_x (p_y)$ particles.  Then, following a
Schwinger boson approach \cite{auerbach11}, the corresponding spin 1/2
operators are defined : $S_{z,{\bf r}} = (n_{\uparrow,{\bf r}} -
n_{\downarrow,{\bf r}})/2$, $S_{+,{\bf r}} = b^\dagger_{\uparrow,{\bf
    r}}b_{\downarrow,{\bf r}}$, and $S_{-,{\bf r}} =
b^\dagger_{\downarrow,{\bf r}}b_{\uparrow,{\bf r}}$.  With these
definitions, the model is rewritten, up to some constants, as
\begin{eqnarray}
H_{1/2} &=& -t \sum_{{\bf r}}\sum_{\hat\alpha=\hat x,\hat y}
\sum_{\sigma=\uparrow, \downarrow}\left(b^\dagger_{\sigma, \bf r}
b_{\sigma, {\bf r}+\hat{\alpha}} + {\rm h.c.}\right) \\ &&+\sum_{\bf
  r} \left( U \frac{n_{\bf r}(n_{\bf r}-1)}{2} - \frac{2U}{3}S_{x,{\bf
    r}}^2+ 2\Delta S_{z,{\bf r}} \right) \nonumber
\end{eqnarray}
where $S_{x,{\bf r}}=(S_{+,{\bf r}}+S_{-,{\bf r}})/2$ and plays a
r\^ole similar to the $L_{z,{\bf r}}$ operator in the original model.

Up to the external field term along the $z$-direction, this is the
model that was studied in \cite{deforges11,deforges12} with $U_0=U$
and $U_2=-U/3$ for the values of the parameters used in these
articles.  For $\Delta=0$, it was shown that the system is always
ferromagnetic at low temperature. On a given site, the absolute value
of the projection of the spin along the $x$ axis is maximized due to
the negative $S_{x,{\bf r}}^2$ term.  The hopping of the particles
then creates an effective ferromagnetic coupling between spins located
on different sites.  This ferromagnetic order is of the Ising class
because of the anisotropy introduced by the $S_{x,{\bf r}}^2$ term. It
is measured by calculating the spin-spin correlation function along
the $x$ axis, related to the correlation of angular momenta $C_z({\bf
  R})$ in the original model (${\bf R} = (R_x,R_y)$),
\begin{equation}
  C_z({\bf R}) = \left\langle L_{z,{\bf r}}L_{z,{\bf r+R}} \right\rangle = 
  4(-1)^{R_x+R_y}\left\langle S_{x,{\bf r}}S_{x,{\bf r+R}} \right\rangle 
\end{equation}
That is, the observed ferromagnetism in the spin$-1/2$ model
corresponds to the staggered angular momentum predicted in the
$p$-band model.

Although it always adopts ferromagnetic behavior at $\Delta=0$, the
system can be in different incompressible Mott phases, at integer
densities and large enough interaction $U$, or in a superfluid phase,
for non integer densities and for integer densities at low enough
interaction $U$.  A constant value, at long distance, of the one
particle Green functions $G_{\sigma}({\bf R}) = \langle
b^{\phantom{\dagger}}_{\sigma,{\bf r}}\, b^\dagger_{\sigma,{\bf
    r+R}}\rangle$ shows that the particles have condensed and that the
system is superfluid.  In terms of $p$-band particles, the Green
functions $G_{x(y)}({\bf R}) =\langle p^{\phantom{\dagger}}_{x(y),{\bf
    r}} p^\dagger_{x(y),{\bf r+R}}\rangle$ have the following
expressions,
\begin{equation}
G_x({\bf R}) = (-1)^{R_x}\, G_\uparrow({\bf R}), \ 
G_y({\bf R}) = (-1)^{R_y}\, G_\downarrow({\bf R})
\end{equation}
That is, a superfluid/BEC phase for spin $1/2$ particles translates
directly into the BEC at non zero momentum phase for the $p$-band
particles, because of the real space phase factors $(-1)^{R_x}$ and
$(-1)^{R_y}$.

The same correspondence holds for the Fourier transforms of these
functions, namely the spin $1/2$ magnetic structure factor at ${\bf
  k}=(0,0)$ becoming an angular momentum structure factor at ${\bf
  k}=(\pi,\pi)$ (we will call this quantity $S_{\rm SAM}$ in the
following) and the density of condensed spin $1/2$ particles at ${\bf
  k}=(0,0)$ becoming the density of condensed $p_x$ and $p_y$
particles at ${\bf k}=(\pi,0)$ and ${\bf k}=(0,\pi)$, respectively,
which will be denoted $\rho_{cx}$ and $\rho_{cy}$ in the following.

Finally, in the spin$-1/2$ model, it is possible to measure the
superfluid density $\rho_s$ through the fluctuations of total winding
numbers \cite{pollock87}
\begin{equation}
\rho_s = \frac{\langle W_x^2 + W_y^2\rangle}{4\beta t}
\end{equation}
At zero temperature, the superfluid density $\rho_s$ and the condensed
densities $\rho_{cx}$ and $\rho_{cy}$ are generally non-zero
simultaneously so we will use $\rho_s$ to determine if we are in the
condensed phase for both models.

\section{Simulation results}

Quantum Monte Carlo simulations using the SGF algorithm
\cite{rousseau08,rousseau08-2} allow us to study the spin$-1/2$ model
at finite temperature for $L\times L$ lattices up to $L=10$ and
inverse temperature up to $\beta t = 80$.  The difficulty of the
simulations is caused by the $S_x^2$ interaction term which changes
the spin projection of the particles. The simulations are performed at
low temperatures in order to obtain the behavior of the ground
states. We will concentrate on $\Delta \ne 0$.

\subsection{Phase diagrams}

\begin{figure}
\includegraphics[width=8.5cm]{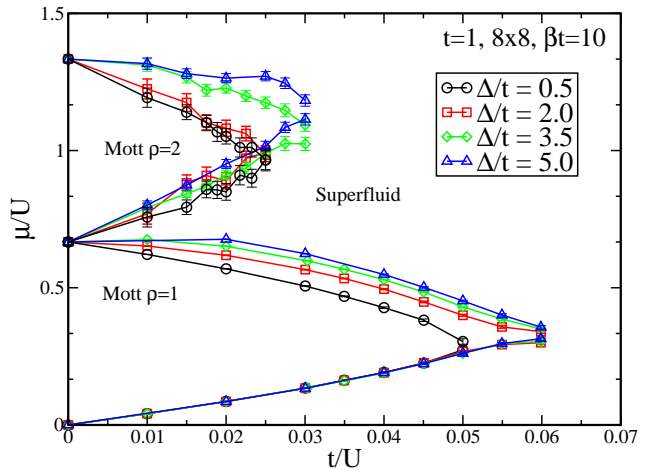}
\caption{(Color online) Phase diagram for $\Delta/t = 0.5, 2.0, 3.5,$ and
  5.0. We observe two Mott phases and a superfluid phase. The extents
of the Mott phases are not varying a lot with $\Delta$, especially
for the $\rho=1$ Mott phase and large values of $\Delta$.
  \label{diagdiffD}
}
\end{figure}

\begin{figure}
\includegraphics[width=8.5cm]{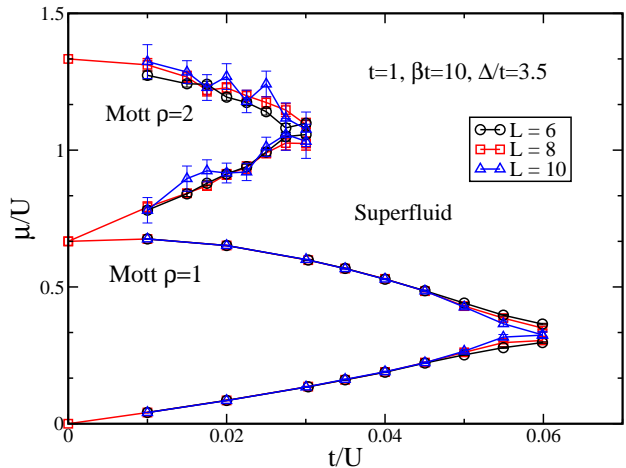}
\caption{(Color online) Phase diagram for $\Delta/t = 3.5$ and
  different sizes. The finite size effects are negligible.
\label{diagD3.5}}
\end{figure}

We expect Mott phases to appear for large enough interactions and
integer densities $\rho$. To determine the extent of these phases in
the $(t/U, \mu/U)$ plane for a given value of $\Delta$ we calculate
the energy $E(N)$ at low temperature for $N=\rho L^2$, $N+1$ and $N-1$
particles. We then determine the boundaries of the Mott lobe as $\mu_+
= E(N+1)-E(N)$ and $\mu_- = E(N)-E(N-1)$, and $\mu_+$ is larger than
$\mu_-$ by a finite value, the charge gap.  As expected, we find Mott
phases occur for integer densities $\rho=1$ and $\rho=2$.  (We did not
go beyond $\rho=2$). We observe that the boundaries of the different
phases are not changed much when $\Delta$ is varied
(Fig.~\ref{diagdiffD}). Since the energy is a local quantity which is
not very sensitive to system size, the boundaries of the Mott lobes
are little changed with system size as seen in Fig.~\ref{diagD3.5}.
We will detail below the nature of these three phases for different
densities.

\subsection{$\rho=1$ case}

\begin{figure}[h]
\includegraphics[width=8.5cm]{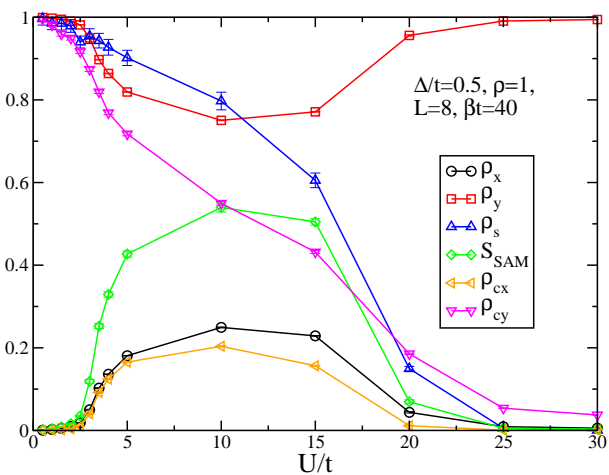}
\caption{(Color online) 
    Cut in the phase diagram at $\rho=1$ and
  $\Delta=0.5$.  The system goes from a Bose condensed phase at low
  interaction to a Mott phase as $\rho_s$ goes to zero with increasing
  interaction $U$.  There is an intermediate SAM-Bose condensed phase.
\label{cutrho1}}
\end{figure}

\begin{figure}[h]
\includegraphics[width=8.5cm]{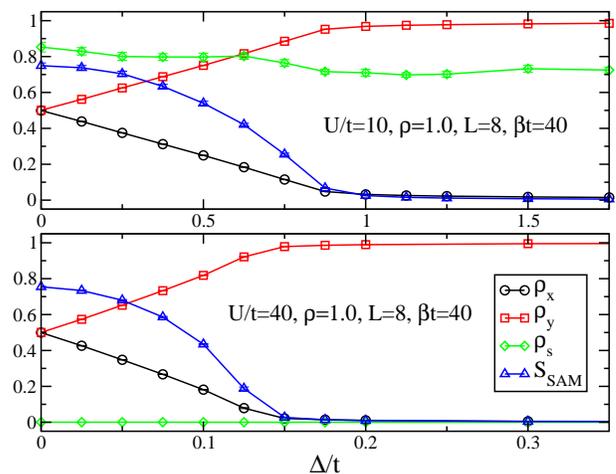}
\caption{(Color online) Cuts in the phase diagram at $\rho=1$
as a function of $\Delta/t$.
 At $U/t=10$ (top), the system is always 
in the superfluid phase ($\rho_s \ne 0$) and we
  observe a staggered angular momentum order when $\Delta < 1 $.
  For $\Delta > 1$, the superfluid is composed of only one dominant
  species. The same happens in the Mott phase for large interaction 
$U/t=40$ (bottom) with a smaller value ($\Delta \simeq 0.15$) for the
disappearance of the SAM order.
\label{cutrho1U40}}
\end{figure}

In Fig.~\ref{cutrho1}, we present the dependence of the particles,
superfluid, and condensate densities and the angular momentum
structure factor on the interaction $U$ at fixed density $\rho=1$ and
$\Delta=0.5$.  We observe, as expected for non zero $\Delta$, that
$\rho_y$ is larger than $\rho_x$. As $U$ is increased, $\rho_x$,
$\rho_y$ and $S_{\rm SAM}$ show a non monotonic behavior. There is a
correlation between $\rho_x$ and $S_{\rm SAM}$ which can be understood
by noting that $S_{\rm SAM}$ order requires a superposition of $p_x$
and $p_y$ particles that is possible only when $\rho_x$ is not zero.
However, the superfluid density decreases monotonically with $U$ and
the system is driven into an incompressible Mott phase where double
occupancy is suppressed at large $U$.  In this Mott phase the model
can then be mapped onto an effective anisotropic spin 1/2 Heisenberg
model. For $\Delta = 0$, the larger coupling of the Heisenberg model
is along the $x$ axis and is equal to $J_x = -9t^2/U$ which leads to
ferromagnetic order.  This is SAM order in terms of $p$-band bosons
\cite{kuklov03,fossfeig11}.  The $\Delta$ term acts as an external
magnetic field along the $z$ axis and tends to destroy the
ferromagnetic/SAM order.  In Fig.~\ref{cutrho1}, the Mott phase is
composed only of $p_y$ particles ($\rho_y \simeq 1$) as $\Delta$ is
large enough to overcome the ordering of effective spins along the
$x$-axis.  There is, then, no SAM order in the Mott phase in this
case.  However, the Mott phase shows a SAM order for small enough
values of $\Delta$ (see Fig.~\ref{cutrho1U40}, bottom).

Returning to the low interaction regime, there are two different
superfluid phases observed in Fig.~\ref{cutrho1}.  At low interaction
$U$, the system is well described as a collection of free particles
and, since the $\Delta$ term lowers the energy of $p_y$ states, the
particles condense in this state. $L_z$ is then on average zero on
each site and no SAM is observed. The $p_y$ bosons form a condensate
and the phase is a striped superfluid because $\rho_{cy}$ is nonzero.
For intermediate interaction, we observe a SAM superfluid where
$\rho_{cx}$, $\rho_{cy}$ and $S_{\rm SAM}$ are non zero at the same
time. For such moderate $U$, the interaction is not large enough to
block particles in a Mott phase but, to lower the energy, the system
adopts states that have non zero $L_z$ on a site and the hopping terms
lock the relative phases into an SAM order.  As in the Mott case, the
SAM order will disappear if the system is composed of only $p_y$
particles due to a large value of $\Delta$ (see Fig. \ref{cutrho1U40},
top).

\begin{figure}[!h]
\includegraphics[width=8.5cm]{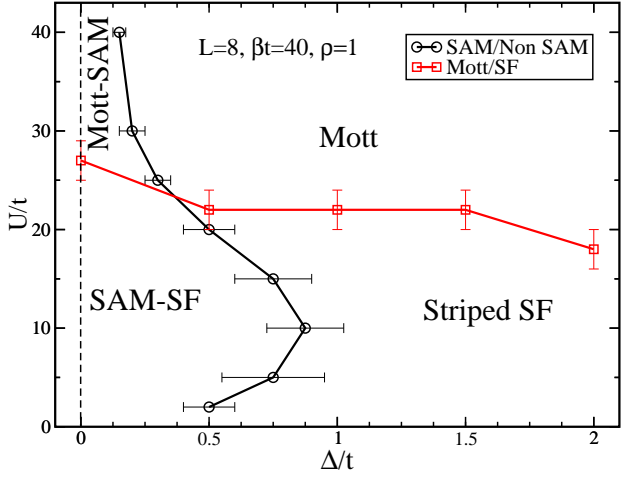}
\caption{(Color online) Phase diagram for $\rho=1$ as a function of
  $U/t$ and $\Delta/t$. There are four different phases: a Mott
  phase, a Mott phase with staggered orbital momentum (SAM) order, a
  superfluid phase with SAM order and a striped superfluid.
  \label{phasediagrho1}}
\end{figure}

Finally, using different cuts in the phase diagram at fixed $\Delta$
or $U$, similar to Figs.~\ref{cutrho1} and \ref{cutrho1U40}, we map
the phase diagram presented in Fig.~\ref{phasediagrho1} for
$\rho=1$. We observe that for the range of values we studied, the SAM
order completely disappears for $\Delta > 1$ and the system is in a
Mott phase for $U>25$.

\subsection{$\rho=2$ case}

\begin{figure}[!h]
\includegraphics[width=8.5cm]{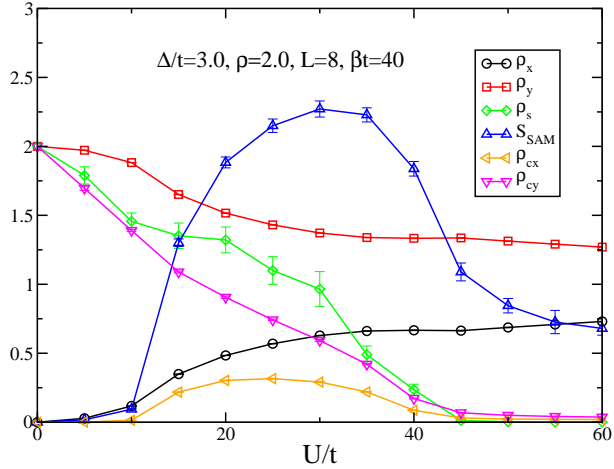}
\caption{(Color online) Cut in the phase diagram at $\rho=2$ as a
  function of $U/t$, the quantities and the observed behavior are
  similar to the $\rho=1$ case (see Fig.~\ref{cutrho1}) but the
  $S_{\rm SAM}$ order is more robust as it is still non zero in the
  Mott phase, despite the much larger value of $\Delta$ ($\Delta/t=3$
  in this case compared to $\Delta/t=0.5$ in Fig.~\ref{cutrho1}).
\label{cutrho2}}
\end{figure}

\begin{figure}[!h]
\includegraphics[width=8.5cm]{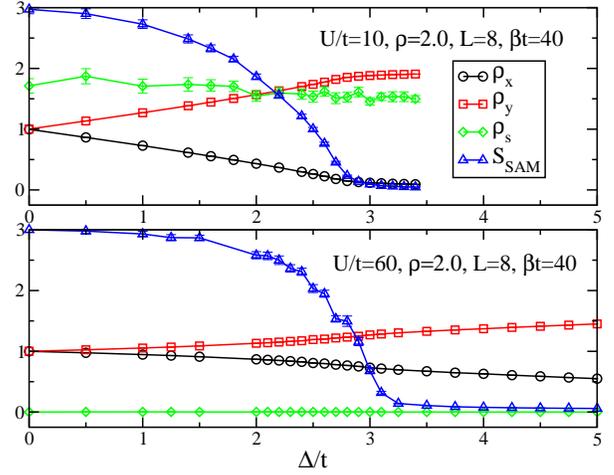}
\caption{(Color online) Cut in the phase diagram at $\rho=2$ as a
  function of $\Delta/t$. The quantities and the observed behavior are
  similar to the $\rho=1$ case (see Fig.~\ref{cutrho1U40}) with a more
  robust SAM order persisting up to $\Delta/t \simeq 3$. 
\label{cutrho2Delta}}
\end{figure}

\begin{figure}[!h]
\includegraphics[width=8.5cm]{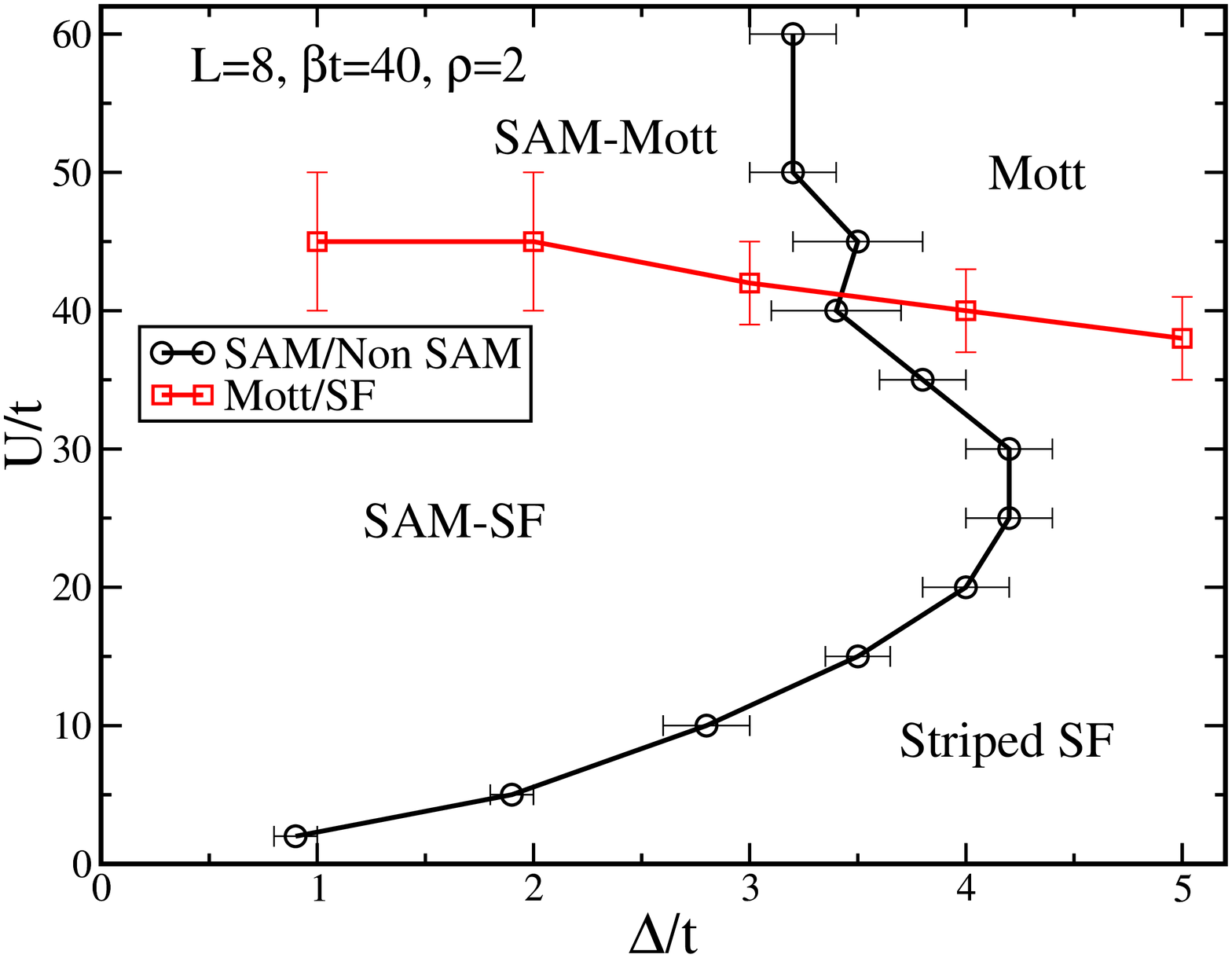}
\caption{(Color online) Phase diagram for $\rho=2$ as a function of
  $U/t$ and $\Delta/t$. The SAM phases are more robust than in the
  $\rho=1$ case (Fig.~\ref{phasediagrho1}) but the same four phases
  are observed.
\label{phasediagrho2}}
\end{figure}

We performed the same analysis for the $\rho=2$ case and found similar
results (see Figs.~\ref{cutrho2}, \ref{cutrho2Delta}, and
\ref{phasediagrho2}).  However, the SAM is much more robust, as it
persists up to $\Delta=4$. This can be understood by recalling that
the $L^2_{z{\bf r}}$ term scales as the square of the density. Thus
the associated SAM coupling of the angular momentum should grow
rapidly with the density and the stripe phase is therefore more
difficult to observe.

\subsection{Non integer densities}

\begin{figure}
\includegraphics[width=8.5cm]{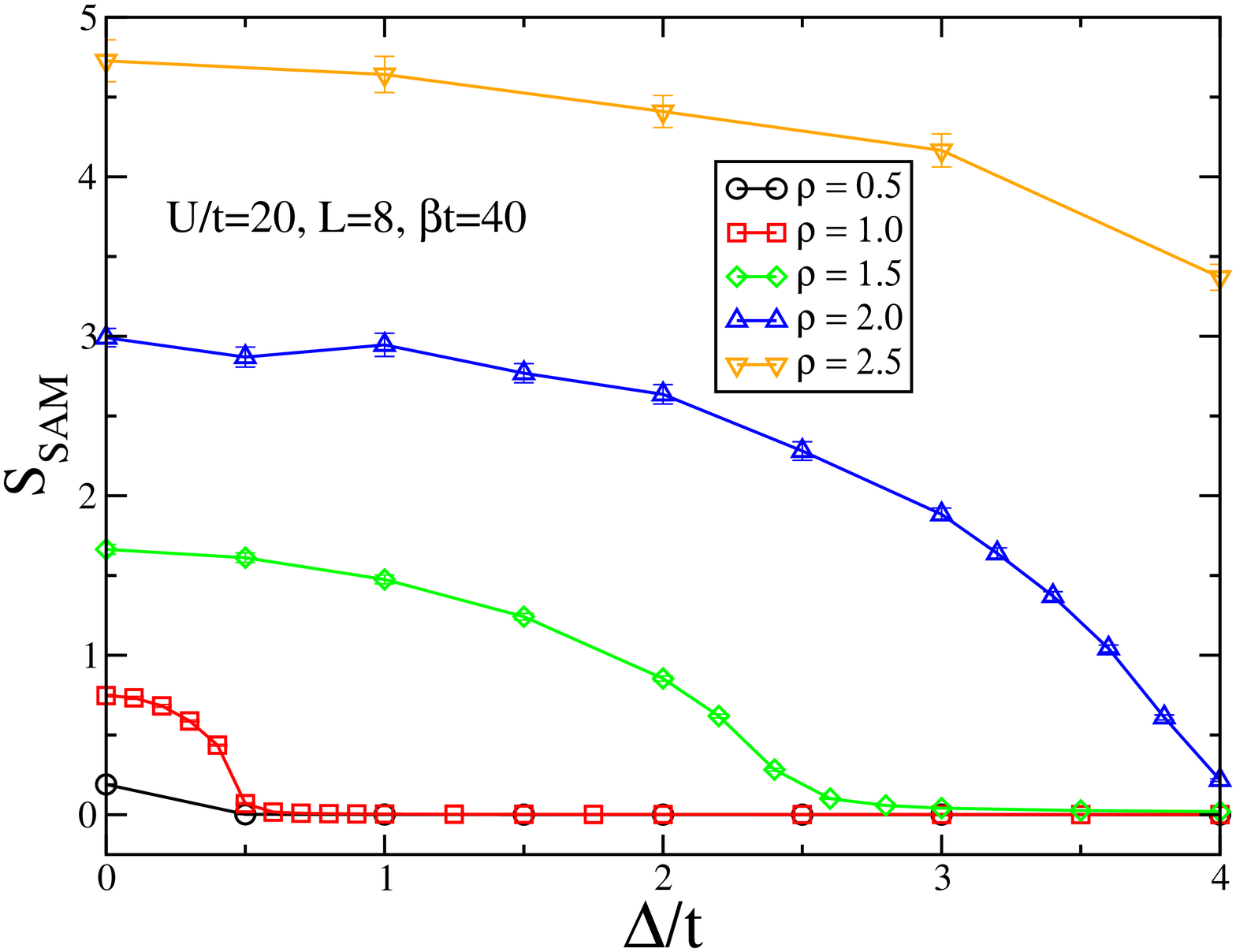}
\caption{ (Color online) Cuts as functions of $\Delta$ for $U=20$ and
  different integer and non integer densities. For these regimes, the
system is always superfluid, even for integer densities. We observe
that, as the density is increased, the structure factor $S_{\rm SAM}$
becomes larger and larger and persists for larger values of
$\Delta$. The SAM order is found for all densities, provided $\Delta$
is small enough.
\label{cutdiffdens}
}
\end{figure}

At non integer densities, the system is of course always superfluid.
As at commensurate fillings $\rho=1$ and $\rho=2$, we observe a
transition between the SAM superfluid phase and the stripe superfluid
as $\Delta$ is increased. The limiting value of $\Delta$ grows with
increasing density, a trend already evident in comparing results for
the two integer fillings (Fig.~\ref{cutdiffdens}). The value of
$\Delta$ at which the SAM disappears corresponds to the system being
populated almost entirely by $p_y$ particles.  We observe the SAM
order for any density, which is in contradistinction with the results
from \cite{wu09} where the SAM phase was expected to appear only for
$\rho \ge 2$.

\section{Conclusion}

We studied a model for $p$-band superfluidity \cite{liu06,wu09} using
exact quantum Monte Carlo simulations. We found that the phase diagram
of the model is composed of a striped superfluid, a superfluid with
simultaneous staggered angular momentum order (SAM), a SAM Mott and a
Mott phase.  Contrary to what was expected, \cite{liu06,wu09} our QMC
simulations indicate that the SAM superfluid phase can be observed for
any density, and not just for densities larger than two, although it
becomes more robust as the density is increased. The presence of the
SAM order in the $\rho=1$ Mott can be understood with an analysis in
terms of an effective Heisenberg model.

Despite its differences with the experiment of Hemmerich and
co-workers (namely the absence of the $s$-wave sites), the simple
model studied here gives a good qualitative description of the
results. It reproduces the two observed superfluid phases, the SAM and
stripe superfluids, with the same condensations at non zero momenta.
It also reproduces the transition between these two exotic superfluids
driven by the energy difference $\Delta$ between the two species.
Moreover, our results suggest that in a system with stronger repulsive
interaction, a similar phase transition between a SAM and a non-SAM
phase could be observed in a Mott insulating phase.

Here, we focused exclusively on the isotropic case,
$t_{\perp}=t_{\parallel}$, although an experimental realisation would
have anisotropies, $t_{\perp}\neq t_{\parallel}$.  We have done some
preliminary simulations (not shown here) for $t_{\perp}$ and
$t_{\parallel}$ values that are not very different.  These indicate
that the physics remains qualitatively the same.  The case of extreme
anisotropy, $t_{\perp}/t_{\parallel} \ll 1$ and even $t_{\perp}=0$
does differ \cite{isacsson05} and merits special attention.

\acknowledgements 

We would like to thank L. de Forge de Parny for stimulating
discussions.  This work was supported by: the CNRS-UC Davis EPOCAL LIA
joint research grant; by NSF grant OISE-0952300; and ARO Award
W911NF0710576 with funds from the DARPA OLE Program. C. W.  is
supported by NSF DMR-1105945 and AFOSR FA9550-11-1-0067(YIP).

\end{document}